\documentclass[aps,prb,reprint,longbibliography,showkeys,superscriptaddress]{revtex4-1}
\usepackage{bm,graphicx,tabularx,array,booktabs,dcolumn,xcolor,microtype,multirow,amscd,amsmath,amssymb,amsfonts,physics,tikz,wrapfig,enumitem,float}

\usepackage{tikz}
\usetikzlibrary{arrows.meta,positioning,shapes.misc,arrows}

\usepackage[version=4]{mhchem}
\usepackage[utf8]{inputenc}
\usepackage[T1]{fontenc}
\usepackage[normalem]{ulem}
\usepackage{times}

\usepackage{goldstone}
\usepackage{feynmf}

\usepackage{setspace}
\usepackage{}
\usepackage[]{graphicx}

\usepackage{siunitx}
\usepackage{fixltx2e}
\DeclareSIUnit\hartree{E\textsubscript{h}}

\usepackage{hyperref}
\hypersetup{
    colorlinks=true,
    linkcolor=blue,
    filecolor=blue,      
    urlcolor=blue,	
    citecolor=blue
}

\usepackage{titlesec}
\titleformat{\section}{\bfseries}{}{0em}{}
\titleformat{\subsection}{\it\bfseries}{}{0em}{}
\titleformat{\subsubsection}{\it}{}{0em}{}
\titlespacing\section{0pt}{10pt plus 2pt minus 2pt}{2pt plus 2pt minus 2pt}
\titlespacing\subsection{0pt}{10pt plus 2pt minus 2pt}{2pt plus 2pt minus 2pt}
\titlespacing\subsubsection{0pt}{10pt plus 2pt minus 2pt}{2pt plus 2pt minus 2pt}

\definecolor{hughgreen}{RGB}{127, 00, 255}

\newcommand{\blue}[1]{\textcolor{blue}{#1}}

\usepackage{subcaption}
\usepackage{cleveref}

\newcommand{\foreign}[1]{\textit{#1}}

\newcommand{\ansatz}{\foreign{ansatz}}
\newcommand{\ansatze}{\foreign{ans\"{a}tze}}

\newcommand{\e}{\mathrm{e}}
\renewcommand{\i}{\mathrm{i}}

\newcommand{\hkap}{\hat{\kappa}}

\newcommand{\hH}{\hat{H}}
\newcommand{\hA}{\hat{A}}
\newcommand{\hB}{\hat{B}}

\newcommand{\SuppI}{supplementary materials}

\newcommand{\UCAM}{Yusuf Hamied Department of Chemistry, University of Cambridge, Lensfield Road, Cambridge, CB2 1EW, U.K.}
\newcommand{\UOX}{Physical and Theoretical Chemistry Laboratory, University of Oxford, South Parks Road, Oxford, OX1 3QZ, U.K.}

\begin{document}

\setlength{\abovedisplayskip}{7pt}
\setlength{\belowdisplayskip}{7pt}

\title{Exact electronic states with shallow quantum circuits through global optimisation}

\author{Hugh~G.~A.~Burton}
\email{hgaburton@gmail.com}
\affiliation{\UOX}
\author{Daniel~Marti-Dafcik}
\affiliation{\UOX}
\author{David~P.~Tew}
\affiliation{\UOX}
\author{David~J.~Wales}
\affiliation{\UCAM}

\date{\today}

\begin{abstract}
Quantum computers promise to revolutionise electronic simulations by
overcoming the exponential scaling of many-electron problems.
While electronic wave functions can be represented using a product of
fermionic unitary operators, shallow quantum circuits for exact states have not yet been achieved.
We construct universal wave functions from gate-efficient, symmetry-preserving
fermionic operators by 
introducing an algorithm that globally optimises the wave function
in the discrete ansatz design and the continuous parameter spaces.
Our approach maximises the accuracy that can be obtained 
with near-term quantum circuits. 
Highly accurate numerical simulations on strongly correlated molecules,
including water and molecular nitrogen, and the condensed-matter Hubbard
model, demonstrate that our algorithm reliably advances the state-of-the-art,
defining a new paradigm for quantum simulations featuring strong electron correlation.
\end{abstract}

\maketitle

%
\raggedbottom
Computing molecular properties and processes relies on 
solving the Schr\"odinger equation to find the many-electron wave function.
However, this problem scales exponentially with 
the number of electrons and practical digital algorithms rely on approximations that fail for 
strong electronic correlation. 
Therefore, accurate simulations of technologically important processes, such as transition metal 
catalysis and high-temperature superconductivity,\cite{Bauer2020} remain challenging.
Gate-based quantum computation offers a revolutionary approach with the potential to overcome 
this exponential scaling and provide a general solution to quantum chemistry.\cite{AspuruGuzik2005}

Current and near-term quantum devices are impaired by noise and are limited to 
shallow quantum circuits.\cite{McArdle2020}
The most promising near-term approaches optimise the parameters of an \ansatz{} 
for the quantum state, and algorithms such as the variational quantum 
eigensolver\cite{Peruzzo2014} (VQE) have the potential to outperform classical computational 
methods with modest quantum resources.
However, the best choice of \ansatz{} for strongly correlated wave functions is far from clear, 
raising the question of whether near-term quantum computing will be useful 
for addressing unsolved quantum chemical problems.

We present a method for designing highly compact
\ansatze{} for quantum computational chemistry by combining products of symmetry-preserving 
fermionic operators with a global optimisation algorithm, 
the \textit{Discretely Optimised	Variational Quantum Eigensolver} (DISCO-VQE), 
which optimises both the discrete sequence of operators and their variational parameters.
Numerical simulations using this approach demonstrate that 
accurate wave functions for strongly correlated electronic systems can be parametrised
with efficient quantum circuits. 
These advances provide a route towards practical simulations of strong electron 
correlation on quantum devices, bringing us closer to solving 
challenging problems in quantum chemistry.

\section{Unitary product states} 

All \ansatz{}-based approaches apply a unitary transformation
$\hat{U}$ to an initial state $\ket{\Phi_0}$ in the form of a quantum circuit, 
constructed through a series of parametrised building blocks as 
$\hat{U} = \hat{U}_1(\theta_1) \cdots \hat{U}_N(\theta_N)$. 
The parameters $\theta_i$ are optimised on a classical computer by extremising
an objective function evaluated using the quantum computer.
This product of unitary transformations defines a family of wave functions that we 
term unitary product states (UPS). 
The UPS paradigm is completely general and encompasses every \ansatz{} that has 
been proposed for quantum computational chemistry so far.

The predominant \ansatze{} use either ``hardware-efficient'' unitary operators constructed from qubit 
operations that minimise the depth of a specific quantum circuit,\cite{Kandala2017} 
or ``physically-motivated'' fermionic unitary operations\cite{Anand2022} based on the 
second-quantised representation of electronic systems.\cite{HelgakerBook}
Second-quantised fermionic operators ensure that the quantum circuit satisfies particle 
number symmetry and Pauli antisymmetry, making them appropriate for electronic states.
While physically-motivated operators typically reduce the number of \ansatz{} parameters, 
each fermionic operator requires multiple gate operations that lead to deep quantum circuits. 
Therefore, there is an apparent trade-off between symmetry-preservation and gate 
efficiency.\cite{Anand2022}
Conserving physical symmetries 
ensures that truncated wave functions predict accurate properties 
beyond the energy and can be used as
initial states for fault-tolerant quantum phase estimation, where efficiency 
depends on the overlap of the initial and exact states.

Initial developments using fermionic operators were based on Unitary Coupled 
Cluster\cite{Peruzzo2014,Anand2022} (UCC) schemes, inspired by the success of 
many-body Coupled Cluster theory for classical computing.\cite{Bartlett2007}
The UCC unitary transformation is an exponential of a sum of anti-Hermitian fermionic operators, 
usually restricted to one- and two-body excitations.\cite{Anand2022}
The first step to represent the UCC \ansatz{} as a quantum circuit is converting this
operator into the UPS form. 
Since fermionic operators generally do not commute, expanding the sum requires a 
Trotter approximation that is only exact in the infinite limit:
$\e^{\hA+\hB} = \lim_{m\rightarrow\infty}(\e^{\hA/m}\,\e^{\hB/m})^m$.

Instead, focus has shifted to a particular UPS form of the UCC wave function 
called disentangled UCC.\cite{Evangelista2019} 
While it has been shown theoretically that exact electronic states can be represented
using a suitably-ordered infinite product of one- and two-body fermionic 
operators,\cite{Evangelista2019} in practice, finite expansions are 
sensitive to the operator order 
and often fail for strong correlation.\cite{Izmaylov2020,Grimsley2020}
Moreover, spin symmetry corresponding to the $\hat{S}^2$ operator is not conserved
because the non-commuting spin components of unpaired two-body 
operators appear independently in the unitary product.\cite{Tsuchimochi2020}
The ADAPT-VQE\cite{Grimsley2019} algorithm, and its 
extensions,\cite{Tang2021,Chan2021,Tsuchimochi2022,Yordanov2021}
iteratively build a UPS from a pool of one- and two-body fermionic 
operators, selecting the operator with the largest energy improvement at each step.
Although promising, these approaches are not guaranteed to converge to the 
exact state\cite{Shkolnikov2021,Rubin2021} or identify the most accurate representation 
with the fewest number of operators, and can spontaneously break spin symmetry.

The ideal unitary product \ansatz{} should meet the following requirements:
\begin{itemize}[itemsep=0em,topsep=0.25em]
\item {Fidelity --- symmetries of the electronic Hamiltonian should be preserved by each operator};
\item {Universality --- any electronic eigenstate should be accessible};
\item {Practicality --- the circuit depth and number of entangling gates should be small}.
\end{itemize}
Here, we show that these requirements can be fulfilled by a
symmetry-preserving unitary product state (s-UPS) that conserves
Pauli antisymmetry, particle number symmetry, $\hat{S}_z$ and $\hat{S}^2$ spin symmetry
using a pool of elementary fermionic operators that grows quadratically with the system size.
We prove that this \ansatz{} is universal and demonstrate that it is gate-efficient.
The appropriate choice and ordering of the operators is system-dependent and 
simultaneously optimising the sequence of operators and their variational parameters 
is challenging.
We realise this goal by introducing the DISCO-VQE algorithm,
which finds the best ordered set of operators for a given number of operators.
Our approach yields gate-efficient quantum circuits that prepare an accurate s-UPS for 
strongly correlated systems using remarkably few variational parameters.

\section{Symmetry-preserving unitary product states} 

Fermionic unitary operators are defined through an exponential of 
anti-Hermitian second-quantised operators
$\hkap_{p q \dots}^{r s \dots} 
= 
a_{r}^{\dagger} a_{s}^{\dagger} \cdots  
a_q^{\vphantom{\dagger}} a_p^{\vphantom{\dagger}}- 
a_{p}^{\dagger} a_{q}^{\dagger} \cdots  
a_s^{\vphantom{\dagger}} a_r^{\vphantom{\dagger}}$. 
These operators ensure Pauli antisymmetry, conserve particle number, and can be mapped onto 
qubit operators using fermion-to-qubit encodings.\cite{Jordan1928,Bravyi2002}
We define a symmetry-preserving unitary product state (s-UPS) of length $M$ as
\begin{equation}
	\ket{\Psi(\bm{t},\boldsymbol{\mu})} 
    = 
    \prod_{i=1}^{M} \e^{ t_i\, \hkap_{\mu_i} } \ket{\Phi_0},
	\label{eq:exactWfn}
\end{equation}
where $\ket{\Phi_0}$ is an arbitrary initial state, and only spin-adapted generalized 
one-body and paired two-body anti-Hermitian second-quantised operators appear in the 
product, that is 
$\hkap_{\mu_i} \in  \{\hkap_{p}^q + \hkap_{\bar{p}}^{\bar{q}}, \hkap_{p \bar{p}}^{q \bar{q}}\}$.
The indices $p,q$ denote arbitrary molecular spin orbitals and the absence (presence) 
of an overbar indicates a high-spin (low-spin) orbital.
Individual operators may appear multiple times and each operator has its own
continuous parameters, defined by the continuous coordinates $\bm{t} = \qty(t_1, \dots, t_M)$.
Because these anti-Hermitian operators in general do not commute, the \ansatz{}
depends on the choice and ordering of the unitary operators, denoted
by the ordered set of indices $\boldsymbol{\mu} = \qty(\mu_1, \dots, \mu_M)$.

Using generalised fermionic operators, where $p,q,r,s$ correspond to arbitrary 
molecular spin orbitals, removes any connection between the operators in the s-UPS \ansatz{}
and the separation of occupied or virtual molecular orbitals in some mean-field reference.
In contrast to the many-body perturbation philosophy,
the generalised one- and two-body operators are Lie algebra generators for unitary transformations
in the $D$-dimensional Hilbert space and can rotate any initial state onto 
an arbitrary exact state.
Universality of a UPS \ansatz{} built from all generalised one- and two-body fermionic operators
has already been proved\cite{Evangelista2019,Izmaylov2020} by realising that
the full $\frac{1}{2} D (D-1)$ dimensional Lie algebra\cite{GilmoreLieBook} 
can be generated using operators that can
be decomposed as nested commutators of the generalised one- and two-body fermionic operators.
The universality of the s-UPS \ansatz{} follows from the additional observation that 
generalised two-body operators can be obtained through nested commutators of one-body 
and paired two-body operators (\SuppI{}, section S1).

Since our elementary operators are spin-adapted 
(i.e., the commutators $[\hat S^2,\hkap]$ and $[\hat S_z,\hkap]$ vanish) 
our \ansatz{} conserves the spin quantum numbers of the initial state, avoiding 
spin-constrained optimisation\cite{Tsuchimochi2020} or spin projection.\cite{Tsuchimochi2022} 
Conserving spin symmetry for each operator ensures that
the s-UPS \ansatz{} is spin-pure for every truncation $M$.
In contrast, spin adaptation of unpaired two-body operators is only obtained by 
exponentiating a sum of non-commuting operators, which requires a Trotter expansion to 
be applied on a quantum circuit. 
The spin adaptation is destroyed when this Trotter expansion is approximated.\cite{Tsuchimochi2020}
%
Furthermore, every Pauli-$Z$ operator in the qubit representation 
of the paired two-body operators $\hkap_{p \bar{p}}^{q \bar{q}}$ can be removed by cancellation, 
hence these operators can be implemented with a constant number of CNOT gates.
In contrast, the CNOT count for unpaired two-body operators grows linearly with the number of orbitals (\SuppI{}, section S2.2).


Universality of the s-UPS \ansatz{} is assured because the product of two
unitary operators corresponds to an exponential of a sum of operators through the
Baker--Campbell--Haussdorff expansion\cite{HallBook} $\e^{\hA} \e^{\hB} = \e^{\hA + \hB + \frac{1}{2}[\hA,\hB] + ...}$.
Since these nested commutators generate the full Lie algebra,
combining a sufficient number of them in a suitable order will eventually
represent any unitary transformation in the Hilbert space.
We postulate that $M\le (D-1)$ will allow any initial state to be rotated into an exact state,
avoiding redundant parameters in the exact wave function, 
and provide numerical examples that support this.
In practice, sufficient accuracy is expected with a much smaller value of $M$.
This perspective contrasts with Trotterised UCC,\cite{Grimsley2020,Evangelista2019}
where the unitary product form arises from expanding an
exponential sum of operators as $\e^{\hA+\hB} = \lim_{m\rightarrow\infty}(\e^{\hA/m}\e^{\hB/m})^m$
and $n$-body operators are decomposed into an infinite
unitary product through nested commutators (see Eq.~(24) in Ref.~\onlinecite{Evangelista2019}).

\section{Discretely optimised variational quantum eigensolver}

While electronic states can be represented using
a product of spin-adapted generalised one- and two-body exponential operators,
this construction requires the correct ordered set of operators.
Fixed operator orderings are typically a poor approximation to the optimal unitary product state, 
while simply increasing the number of operators ultimately leads to redundancies.\cite{Grimsley2019,Lee2019c}
In the absence of \textit{a priori} information,
the optimal operator ordering can be approached as a 
coupled discrete and continuous optimisation problem.
Here, we develop the first algorithm that can globally optimise 
the s-UPS \ansatz{} for a given number of operators. 

Exploiting the variational principle, the optimal s-UPS representation can be identified by minimising the energy
\begin{align}
E(\bm{t},\boldsymbol{\mu}) = 
\mel*{\Phi_0}{ \qty( \prod_{i=M}^{1} \e^{-t_i \hkap_{\mu_i}} ) \hH  \qty(\prod_{i=1}^{M} \e^{t_i \hkap_{\mu_i}} )}{\Phi_0}.
\end{align}
In practice, the non-linearity of the energy with respect to $\bm{t}$ creates a 
non-convex energy landscape with local minima that hinder global 
optimisation.\cite{Grimsley2020,Grimsley2022}
Finding the optimal representation requires a combination of 
continuous optimisation for the operator amplitudes $\bm{t}$ and discrete 
optimisation of the ordered operator set $\boldsymbol{\mu}$.
We have developed a new algorithm,  the 
\textit{Discretely Optimised Variational Quantum Eigensolver} (DISCO-VQE), to tackle the
simultaneous global optimisation problem 
in the continuous and discrete spaces for the s-UPS \ansatz{}.

\begin{figure}[t!]
\includegraphics[width=\linewidth]{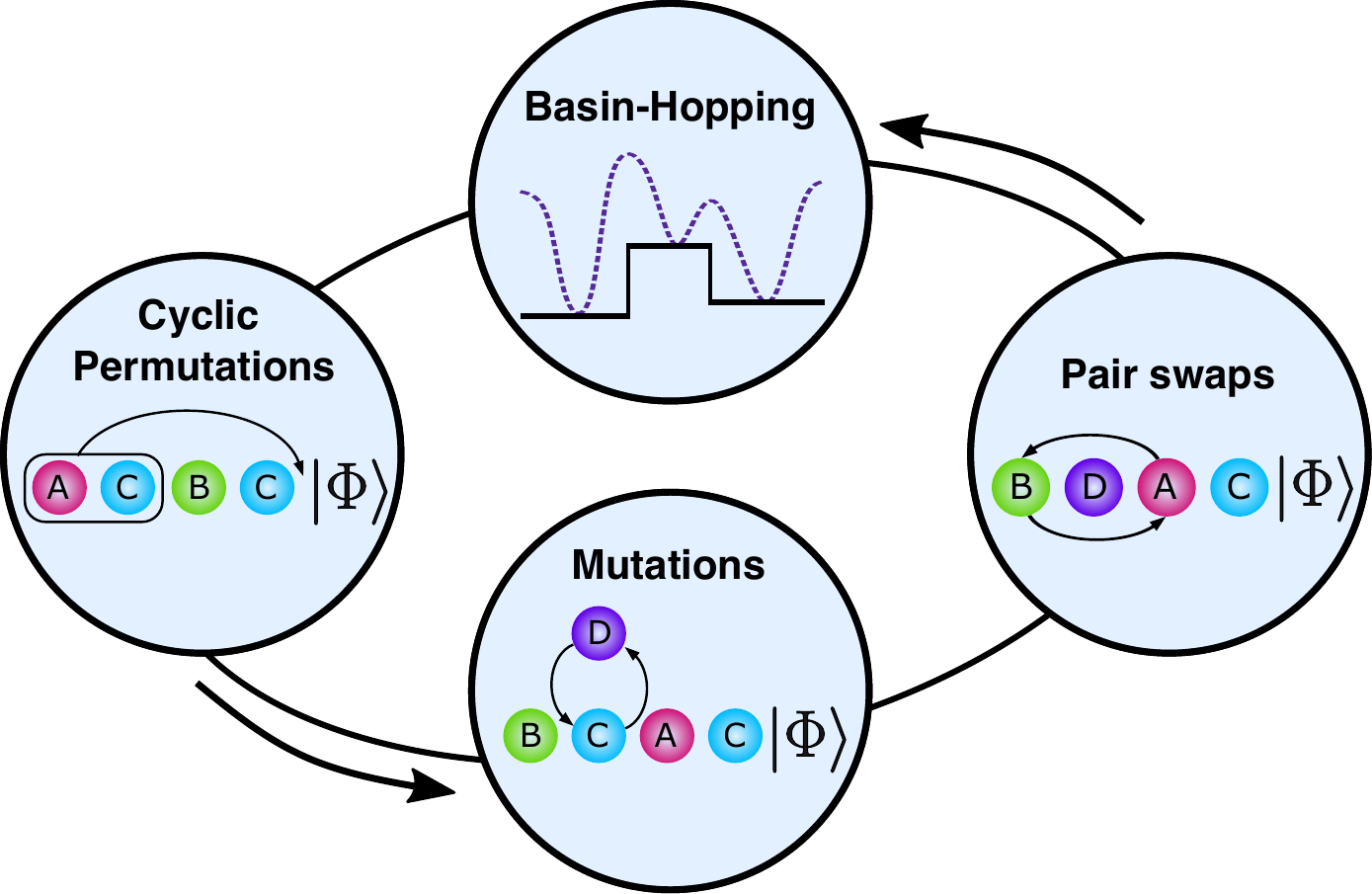}
\caption{
\small
DISCO-VQE is a generalised basin-hopping approach\cite{Schebarchov2013,Schebarchov2014} that optimises the
continuous coordinates $\bm{t}$ and the discrete space of operator orders $\bm{\mu}$
to identify the best s-UPS wave function using cyclic permutations, mutations, and pair swaps of the  unitary operators.
\label{fig:diagrams}
}
\end{figure}

For a given set of operators, we define a local discrete neighbourhood
as the operator sets that can be reached by:
\begin{itemize}[itemsep=0em,topsep=0.25em]
\item {performing a cyclic permutation of the ordered set;}
\item {mutating an operator into another from the pool;}
\item {swapping the position of two operators.}
\end{itemize}
Following the generalised basin-hopping (GBH) approach of
Ref.~\onlinecite{Schebarchov2014}, we define a \textit{biminimum} as
a configuration $(\bm{t}, \boldsymbol{\mu})$ that is a minimum in the
continuous parameters and lies lower in energy than any neighbouring
minimum after a step in the discrete space, including relaxation of the continuous parameters.
Analogous biminima arise in the optimisation of multicomponent
materials,\cite{Schebarchov2013,Schebarchov2014} and for
proteins and nucleic acids when the sequence of amino acids or nucleotides is mutated.\cite{Roeder2018}

To identify biminima, DISCO-VQE alternates between a series of basin-hopping\cite{lis87,Wales1997} steps 
that explore the continuous energy landscape for a
fixed set of operators, and discrete steps in the combinatorial space of
ordered operator sets (Fig.~\blue{\ref{fig:diagrams}}).
Basin-hopping uses random perturbations followed by minimisation to step between local minima
on the continuous energy 
landscape\cite{lis87,Wales1997}
and hence search for the global minimum (\SuppI{}, section S3.1)
After a series of basin-hopping steps, DISCO-VQE identifies the lowest-energy 
configuration that can be reached by performing a 
discrete cyclic permutation followed by reoptimisation of the continuous coordinates.
Similarly, for operator mutations and pair swaps, we identify the lowest energy
mutation or pair swap for each operator and accept this step if it lowers the
energy (\SuppI{}, section S3.2).
By testing mutations and pair swaps for every operator individually, DISCO-VQE 
can completely change the discrete parameters in each macro-cycle and, in principle, 
the optimal ordered operator set could be discovered after one macro iteration.
In practice, the combined space of continuous and discrete
parameters also exhibits many local biminima and uphill moves are required 
to escape from energy traps (\SuppI{}, section S3.3).
The full DISCO-VQE algorithm is detailed in the \SuppI{}, section S3.

\section{Exact wave functions using DISCO-VQE}

\begin{figure*}[t!]
	\includegraphics[width=\linewidth]{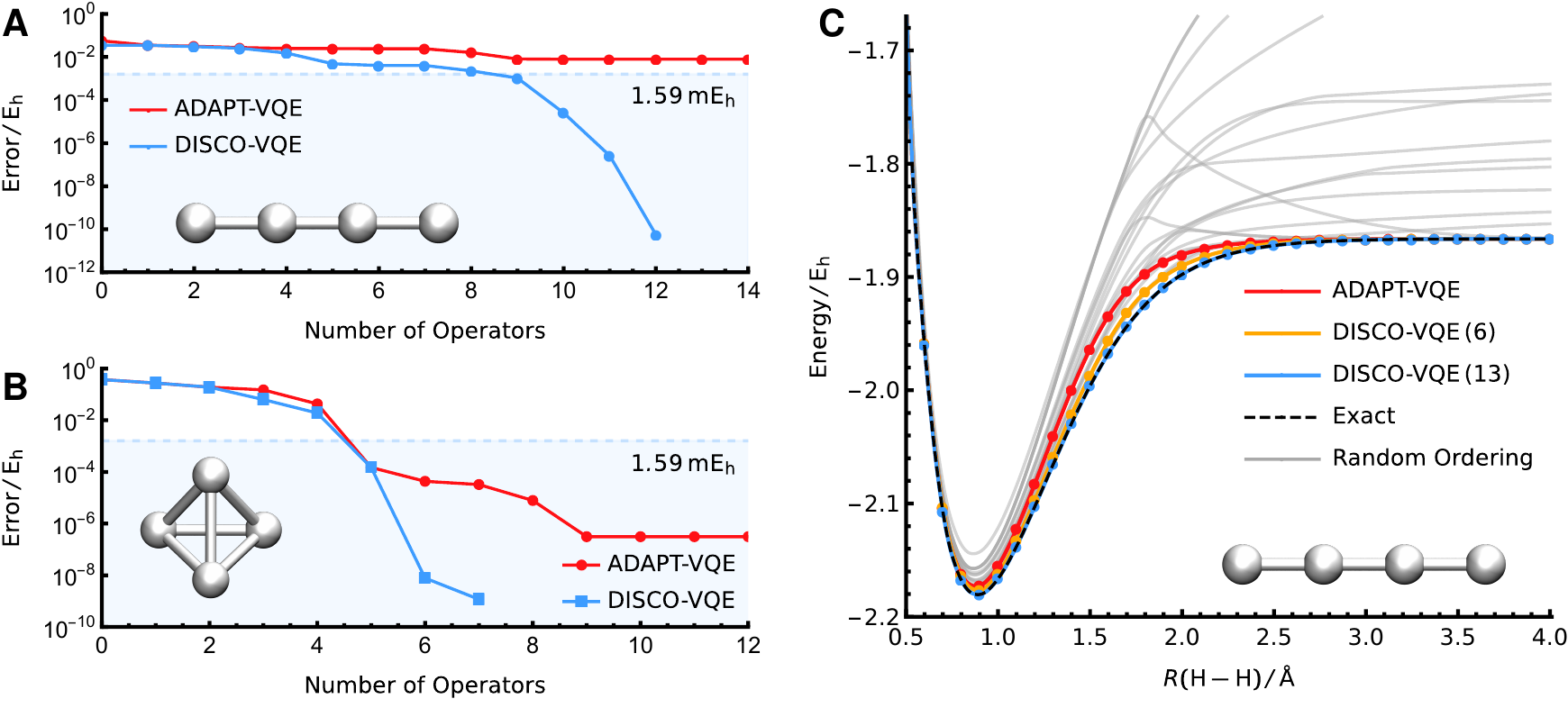}
	\caption{%
\small
		DISCO-VQE produces exact wave functions using global optimisation.
		The panels illustrate the
		convergence of DISCO-VQE and ADAPT-VQE with respect to the number of operators for 
		\textsf{(\textbf{A})} linear \ce{H4}, $R(\ce{H-H}) = \SI{0.90}{\angstrom}$, and 
		\textsf{(\textbf{B})} tetrahedral \ce{H4}, $R(\ce{H-H}) = \SI{1.98}{\angstrom}$,
        using the STO-3G basis set.\cite{Hehre1969} 
		\textsf{(\textbf{C})} DISCO-VQE identifies the exact wave function throughout the linear \ce{H4} binding 
		curve, while ADAPT-VQE and random ordered sets of thirteen operators exhibit a large variation in accuracy.}
	\label{fig:h4}
\end{figure*}

The universality of the s-UPS \ansatz{} means that any exact wave function with an 
arbitrarily small energy error can be obtained using a finite number of unitary operators
selected from the pool of spin-adapted one-body and paired two-body operators.
DISCO-VQE calculations  on the linear and tetrahedral structures of \ce{H4} 
(\SuppI{}, section S4) show that chemical accuracy ($\SI{1.59}{\milli\hartree}$) 
can be obtained with only nine and five operators, respectively
(Fig.~\ref{fig:h4}\textcolor{blue}{\textbf{A--B}}).
Furthermore, ground states with an energy error within the precision of 
our numerical simulations ($10^{-9}\,\mathrm{E_h}$) can be identified using thirteen (linear) or 
eight (tetrahedron) operators.
These results support our postulate that an exact s-UPS representation can be 
constructed with a finite number of operators less than the Hilbert space size and show that 
the spin-adapted one-body and paired two-body operators are sufficient for defining a universal \ansatz{}.

Global optimisation of the continuous coordinates and discrete operator
ordering is essential for realising the universality of the s-UPS \ansatz{}.
The algorithmic advance provided by DISCO-VQE is clear from comparison with 
ADAPT-VQE calculations using the same pool of operators.
By adding one operator at a time, ADAPT-VQE performs a local optimisation 
that can converge onto a non-exact wave function, as demonstrated 
in Fig.~\ref{fig:h4}\textcolor{blue}{\textbf{A--B}}.
This stagnation of the wave function is reminiscent of the
symmetry roadblock described in Ref.~\onlinecite{Shkolnikov2021}
and indicates that ADAPT-VQE can struggle to create the cooperative operator interactions 
required to represent an exact state.
In contrast, DISCO-VQE can escape these local minima in the 
discrete space and provides the first practical algorithm for globally 
optimising the s-UPS wave function.

Computational chemistry requires wave functions that provide consistent
accuracy for every geometry of a molecule.
DISCO-VQE achieves this task for linear \ce{H4} by producing  an exact wave function using 
thirteen operators at every bond length (Fig.~\ref{fig:h4}\textcolor{blue}{\textbf{C}}).
Fixing the operators to the ordering identified for the exact solution at equilibrium 
gives near-exact energies across the full binding curve, with a maximum error of 
$\SI{0.4}{\micro\hartree}$  (Fig.~\textcolor{blue}{\textbf{S2}}), 
demonstrating that highly-accurate and transferable fixed \ansatze{} can in principle be defined.
In comparison, random fixed \ansatze{} with
thirteen operators are relatively accurate near equilibrium, but exhibit a 
large variation in the dissociation limit (Fig.~\ref{fig:h4}\textcolor{blue}{\textbf{C}}).
Furthermore, ADAPT-VQE calculations provide an accurate description of equilibrium and
dissociation, but have an appreciable error at intermediate geometries, where static and 
dynamic correlation  compete (Fig.~\ref{fig:h4}\textcolor{blue}{\textbf{C}}).
Only six operators are required to surpass the accuracy of ADAPT-VQE using DISCO-VQE, 
while chemically accurate binding curves can be obtained with ten operators (Fig.~\textcolor{blue}{\textbf{S3}}).

\section{Accuracy of truncated approximations}

\begin{figure*}[t!]
\includegraphics[width=\linewidth]{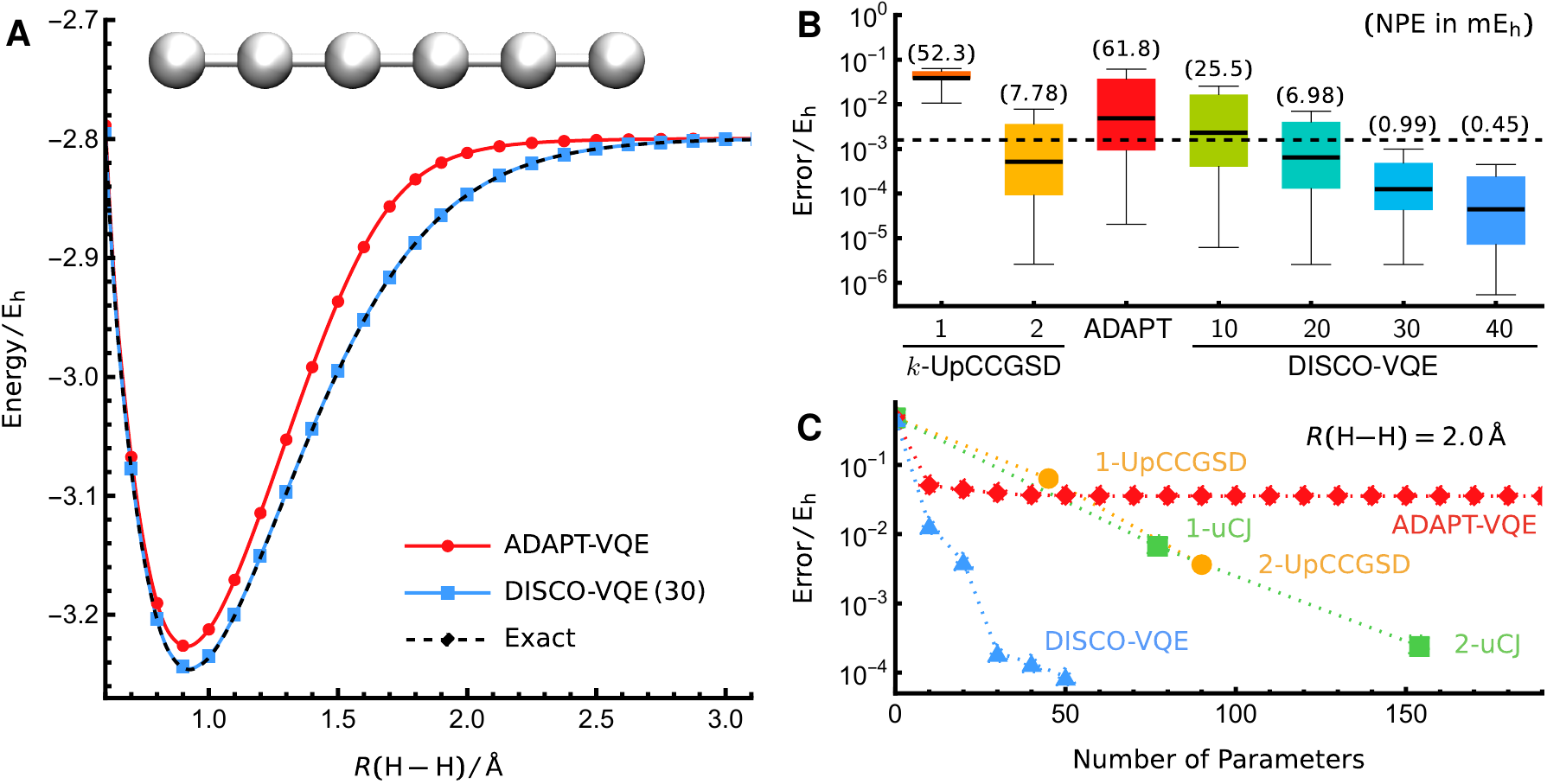}
\caption{
\small \textsf{(\textbf{A})} 
DISCO-VQE  with 30 operators provides chemically accurate energies for the linear \ce{H6} binding curve (STO-3G).
\textsf{(\textbf{B})} 
Comparison of  DISCO-VQE errors and  the previous state-of-the-art methods across the full binding curve.
ADAPT-VQE calculations use the same operator pool.
The un-Trotterised 1-UpCCGSD and 2-UpCCGSD results are taken from Ref.~\onlinecite{Grimsley2020}. 
The non-parallelity error (NPE) is the difference between the maximum and minimum error in the energy.
\textsf{(\textbf{C})} 
DISCO-VQE produces highly accurate energies at $R(\ce{H-H})=\SI{2.0}{\angstrom}$ with
fewer continuous parameters than previous algorithms.
The $k$-uCJ results are taken from Ref.~\onlinecite{Matsuzawa2020}.}
\label{fig:h6}
\end{figure*}

Despite the promise of quantum computers, practical calculations on large molecular systems 
will still require approximate wave function truncations.
Global optimisation of the s-UPS \ansatz{} using DISCO-VQE provides the most 
accurate approximation for a given number of fermionic operators, while conserving spin symmetry 
at every truncation.
For example, DISCO-VQE calculations on strongly-correlated linear \ce{H6} (\SuppI{}, section S5) 
achieve chemical accuracy using only 30 operators (Fig.~\ref{fig:h6}\textcolor{blue}{\textbf{A}}).
In contrast, ADAPT-VQE using the symmetry-pure operator pool stagnates with between 14--40 operators
and fails to identify the exact ground state  (Fig.~\ref{fig:h6}\textcolor{blue}{\textbf{A}}).
DISCO-VQE requires only 10 operators to surpass the accuracy of the ADAPT-VQE 
binding curve (Fig.~\ref{fig:h6}\textcolor{blue}{\textbf{B}}).

Furthermore, DISCO-VQE provides significantly greater accuracy using a smaller number of 
variational parameters than previous related methods with fermionic operators, 
including the un-Trotterised $k$-UpCCGSD\cite{Lee2019c} and the Jastrow-based $k$-uCJ\cite{Matsuzawa2020} 
approaches (Fig.~\ref{fig:h6}\textcolor{blue}{\textbf{C}}).
Only 30 operators are required to surpass the accuracy of the un-Trotterised 2-UpCCGSD wave function
with 90 operators\cite{Grimsley2020} or the 2-uCJ approach with 150 operators,\cite{Matsuzawa2020} 
while the original ADAPT-VQE calculations using non-symmetry-preserving operators required at least
78 operators to obtain comparable results.\cite{Grimsley2019}
Consequently, global optimisation using DISCO-VQE defines a new standard for
the accuracy that can be obtained with a certain number of fermionic unitaries in linear \ce{H6}.

\begin{figure*}[ht!]
    \includegraphics[width=\linewidth]{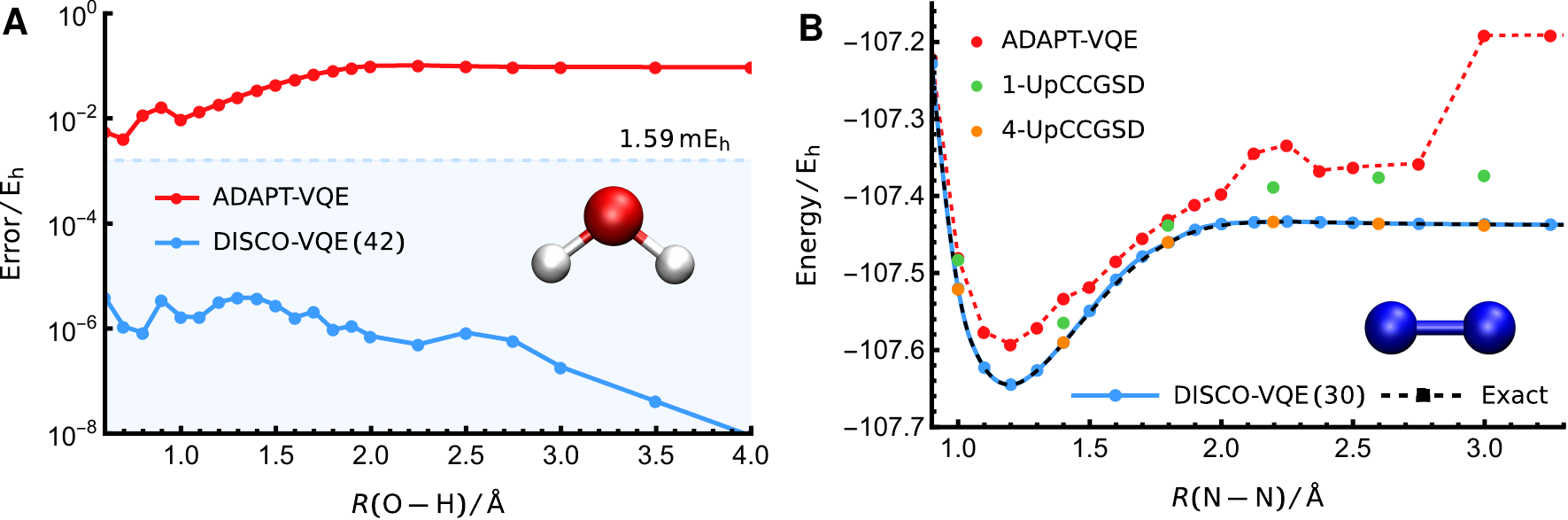}
    \caption{\small Wave functions built using DISCO-VQE accurately describe the potential
        energy surface for strongly correlated molecules.
        The number of operators in these
        DISCO-VQE calculations corresponds to the size of the operator pool in the STO-3G basis set, with
        \textsf{(\textbf{A})} 42 operators for the symmetric
        stretch of \ce{H2O} and \textsf{(\textbf{B})} 30 operators for the dissociation of \ce{N2}.
       The four lowest energy molecular orbitals are frozen in \ce{N2}.
       Un-Trotterised $k$-UpCCGSD results for \ce{N2} are taken from Ref.~\onlinecite{Lee2019c}.}
    \label{fig:h2o_n2}
\end{figure*}

Symmetrically stretched water and the dissociation of \ce{N2} provide important 
testing grounds for the performance of wave function approximations.
DISCO-VQE calculations for a truncated s-UPS \ansatz{} 
produce highly accurate energies for the full \ce{H2O} binding curve (\SuppI{}, section S6), with a non-parallelity 
error\footnote{The non-parallelity error is defined as the difference between the maximum and minimum error along a binding curve.} 
(NPE) of $\SI{0.004}{\milli\hartree}$ (Fig.~\ref{fig:h2o_n2}\textcolor{blue}{\textbf{A}}).
In contrast, ADAPT-VQE, using the same operator pool, fails to identify a chemically accurate
s-UPS wave function at any bond length (Fig.~\ref{fig:h2o_n2}\textcolor{blue}{\textbf{A}}), 
while previous studies using the full pool of one- and two-body operators are much less accurate than 
DISCO-VQE, with an error of  $1.5$ to $\SI{3.0}{\milli\hartree}$.\cite{Sapova2021}
The original un-Trotterised $k$-UpCCGSD \ansatz{} 
required 126 variational parameters to reach an accuracy of $\SI{0.07}{\milli\hartree}$,\cite{Lee2019c}
compared to the 42 operators used for DISCO-VQE,
while recent Trotterised $k$-UpCCGSD calculations using over 250 operators 
have a larger residual error ranging from $0.02$ to $\SI{200}{\milli\hartree}$.\cite{Sim2021} 
Therefore, using DISCO-VQE to globally optimise a s-UPS wave function 
significantly improves the accuracy using fewer fermionic operators than previous state-of-the-art methods.

For the strongly-correlated dissociation of \ce{N2},  DISCO-VQE achieves chemical accuracy 
at nearly every geometry using 30 operators (\SuppI{}, section S7), giving a very 
accurate binding curve (Fig.~\ref{fig:h2o_n2}\textcolor{blue}{\textbf{B}}).
In comparison, previous UPS representations built from randomly ordered sets of all the generalised one- and
two-body operators have residual errors ranging from $16$ to $\SI{64}{\milli\hartree}$,\cite{Grimsley2020}
demonstrating the importance of globally optimising the operator ordering.
The reduction in the number of parameters achieved by DISCO-VQE is demonstrated by comparing to 
the un-Trotterised $k$-UpCCGSD results, which require at least 180 variational parameters to
obtain a better mean-average error\cite{Lee2019c} (Table~\textcolor{blue}{\textbf{S1}}). 
In contrast, ADAPT-VQE calculations with the same operator pool have a significant residual error 
and do not produce a smooth potential energy surface  (Fig.~\ref{fig:h2o_n2}\textcolor{blue}{\textbf{B}}),
while including non-symmetry-conserving one- and two-body operators breaks $\hat{S}^2$ symmetry 
for truncated wave functions.\cite{Tsuchimochi2022}
The s-UPS \ansatz{} conserves the $\hat{S}^2$ symmetry of the initial state 
at every truncation level and global optimisation using DISCO-VQE 
achieves a highly-accurate description of this challenging potential energy surface using a small number of operators.

\section{Maximising the efficiency of quantum circuits}
Applications on near-term quantum hardware must balance the accuracy of an algorithm
 against the  number of two-qubit controlled-NOT (CNOT) gates in the quantum circuit,
which contribute the largest source of hardware noise. 
This challenge has motivated the development of hardware-efficient \ansatze{} 
where the pool of operators can be efficiently represented with a small number of CNOT gates,
such as  qubit-ADAPT-VQE\cite{Tang2021} and qubit-excitation-based (QEB) ADAPT-VQE.\cite{Yordanov2021}
However, these methods do not satisfy fermionic antisymmetry
and may also break  particle number or $\hat{S}_z$ 
symmetry, giving an unphysical electronic state.

\begin{figure}[htb]
	\includegraphics[width=\linewidth]{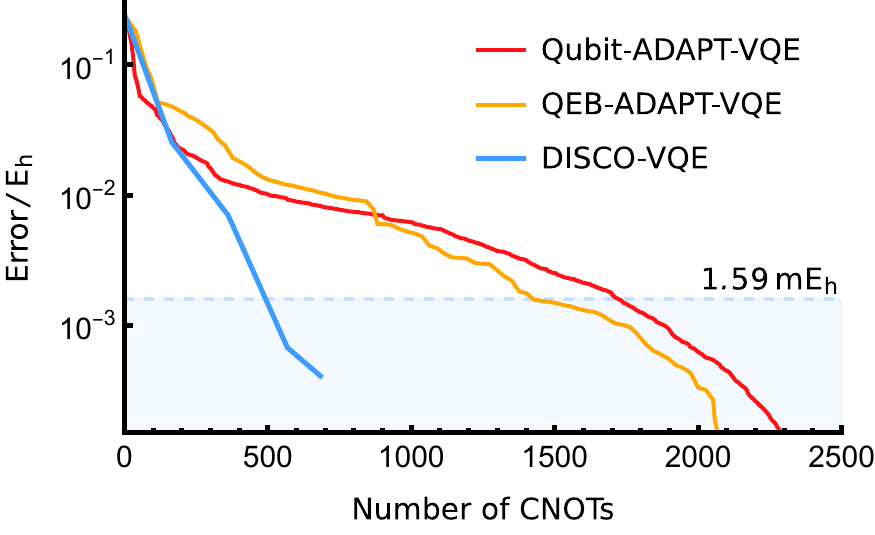}
	\caption{\small DISCO-VQE finds accurate and efficient 
		quantum circuits for linear \ce{H6} at $R(\ce{H-H}) = \SI{1.5}{\angstrom}$
		with significantly fewer CNOT gates than the previous state-of-the-art.
		Data for the qubit-ADAPT-VQE and QEB-ADAPT-VQE methods are taken from Ref.~\onlinecite{Yordanov2021}.
		CNOT counts are evaluated using the efficient circuit design introduced in Ref.~\onlinecite{Yordanov2020} 
        (\SuppI{},  section S2.3).
	}
	\label{fig:cnots}
\end{figure}

Despite employing a fermionic pool of operators, 
DISCO-VQE achieves chemical accuracy for linear \ce{H6} using
less than a third of the CNOT gates  required for previous hardware-efficient \ansatze{} 
(Fig.~\ref{fig:cnots}, see \SuppI{}, section S2.3 for computational details).
These extremely favourable CNOT counts are achieved by removing redundancies from 
the quantum circuit using discrete optimisation
and through the CNOT efficiency of the the paired two-body operators (\SuppI{}, section S2.2).
For the first time, these data demonstrate that efficient quantum circuits can be achieved 
using symmetry-preserving fermionic operators without relying on a hardware-efficient \ansatz{} design, 
creating a new paradigm for quantum-compatible electronic structure methods.

\section{Balancing weak and strong electron correlation}

\begin{figure}[b!]
	\includegraphics[scale=1.0]{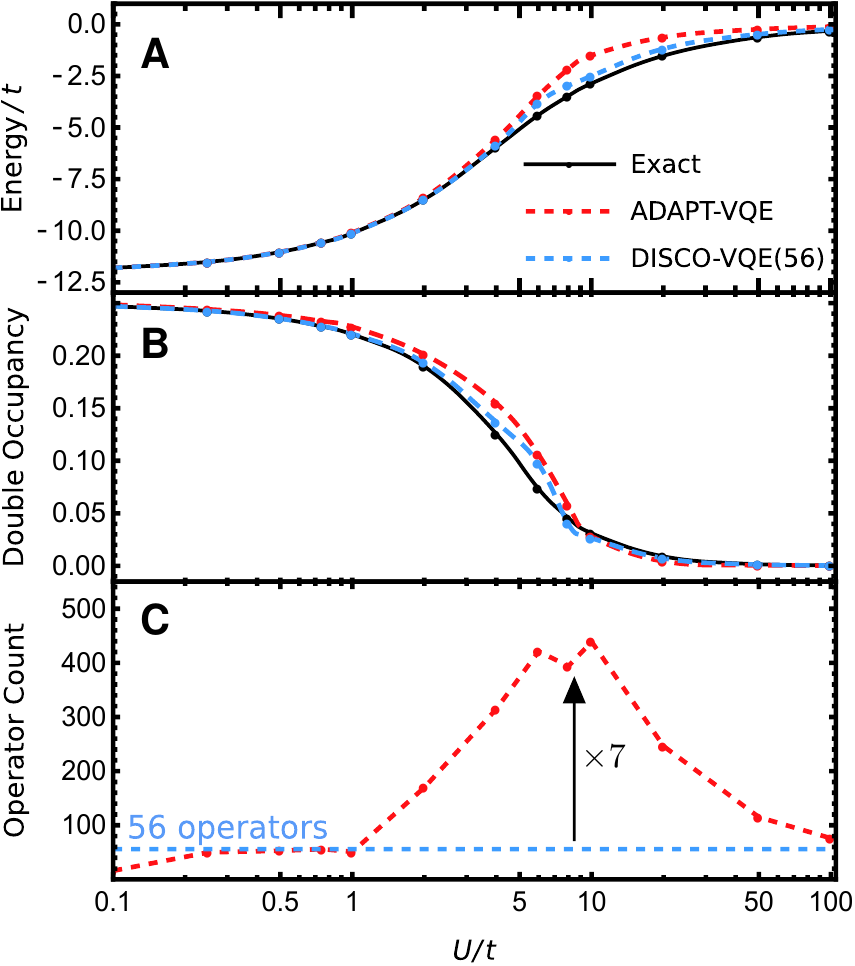}
	\caption{\small DISCO-VQE captures the correct physics for a strongly correlated
		two-dimensional Hubbard lattice ($4\times2$) using 56 operators.
		The DISCO-VQE energy (\textbf{\textsf{A}}) is very accurate in both the weak correlation ($U
		\ll t$) and strong correlation ($U \gg t$) regimes.
		Similarly, DISCO-VQE accurately predicts the double occupancy (\textbf{\textsf{B}}), which becomes exact
		in the weak and strong correlation limits.
		Analogous ADAPT-VQE calculations require seven times more operators at intermediate 
		correlation but give less accurate results (\textbf{\textsf{C}}).
	}
	\label{fig:hubbard_4x2}
\end{figure}

Balanced descriptions of weak and strong electron correlation are essential for quantum chemistry 
but have proved remarkably difficult to achieve in practice.
Many-body perturbation techniques can be quantitatively accurate for weak electron correlation,
but fail to qualitatively describe strongly correlated systems 
where the molecular orbital picture breaks down.
The accuracy of the DISCO-VQE results across the weakly correlated equilibrium geometry and strongly 
correlated dissociation limits of \ce{H6}, \ce{H2O}, and \ce{N2} 
demonstrate that the s-UPS framework can quantitatively predict both types of electron correlation 
using a consistent wave function \ansatze{}.
The accuracy of the s-UPS \ansatz{} for strong correlation highlights the capability to progress beyond the 
many-body framework of coupled cluster theory, and should be considered
as a parametrised unitary rotation in the Hilbert space.

The strongly correlated half-filled Hubbard lattice represents a challenging condensed matter physics
problem for traditional many-body approximations.\cite{LeBlanc2015}
Here, the correlation strength is determined by the balance
of the on-site electron repulsion $U$
and the one-electron hopping term between adjacent sites $t$.
DISCO-VQE calculations with a truncated s-UPS wave function (\SuppI, section S8)
provide excellent accuracy in both the weak correlation ($U\ll t$) and
strong correlation ($U\gg t$) limits (Fig.~\ref{fig:hubbard_4x2}\textcolor{blue}{\textbf{A}}). 
These data highlight the accuracy of the s-UPS \ansatz{} for describing both 
weak and strong correlation using a consistent number of operators and variational parameters.
In comparison, ADAPT-VQE provides a very accurate representation 
of the weakly correlated regime, but deteriorates as the correlation strength increases.
Furthermore, the converged ADAPT-VQE wave function
includes significantly more operators than DISCO-VQE for intermediate correlation ($U/t \approx 10$)
but is less accurate (Fig.~\ref{fig:hubbard_4x2}\textcolor{blue}{\textbf{C}}).
Consequently, global optimisation using DISCO-VQE is essential 
for obtaining quantitatively accurate s-UPS wave functions for both weak and strong correlation
with a consistent quantum resource cost.

Correctly predicting the physical properties of strongly correlated systems 
is just as important as the electronic energy for understanding quantum 
phase behaviour.
The strongly correlated limit of the half-filled Hubbard model is characterised by 
an antiferromagnetic phase where the electrons are unpaired and localised on individual sites.
The double occupancy $\expval*{\hat{n}_{p}\hat{n}_{\bar{p}}}$ measures 
the probability of simultaneously finding two electrons on the same site, which tends to 
zero for $U/t \rightarrow \infty$ and $0.25$ for $U/t \rightarrow 0$.
Truncated s-UPS representations identified using DISCO-VQE accurately predict this change
across different $U/t$ values 
(Fig.~\ref{fig:hubbard_4x2}\textcolor{blue}{\textbf{B}}) and tend to the correct limits.
Therefore, DISCO-VQE accurately predicts both the wave function and the energy, 
providing a physically complete description of strong electron correlation.

\section{Discussion and outlook}

We have shown that an arbitrary electronic state can be parametrised 
with symmetry-preserving unitary product states built from spin-adapted one-body and 
paired two-body fermionic operators.
This \ansatz{} conserves fermionic antisymmetry, particle number symmetry, and 
the quantum numbers corresponding to $\hat{S}^2$ and $\hat{S}_z$ without
any constrained optimisation or spin projection, and 
can parametrise any electronic state.
The flexibility and accuracy of these wave functions comes 
with the additional challenge of identifying the optimal sequence of unitary operators.
DISCO-VQE is the first algorithm that can address
this challenge by performing a coupled discrete and continuous global optimisation of
the wave function.
Numerical simulations demonstrate that our approach delivers
chemically accurate energies for weakly and strongly correlated molecules 
using significantly fewer operators and CNOT gates than previous state-of-the-art techniques.

While the discrete global optimisation in DISCO-VQE increases the computational cost
relative to a fixed \ansatz{} approach,
there are several routes to mitigate this cost in practice.
Firstly, the discrete search over operator mutations or pair swaps 
scales only linearly with respect to the size of the operator 
pool or the number of operators $M$ in the s-UPS \ansatz{}.
Performing individual optimisations for each discrete step in parallel using 
multiple quantum computers would reduce the additional scaling to $\mathcal{O}(M)$. 
Secondly, since the continuous basin-hopping search  represents
a significant proportion of the computational cost, DISCO-VQE will benefit from the 
development of improved quantum numerical optimisation techniques.\cite{Stokes2020,Koczor2022}
Finally, discrete combinatorial optimisation is a rich field and various algorithms
have also been developed for optimising hardware-efficient 
\ansatze{}.\cite{Rattew2019,Chivilikhin2020,Zhang2021,Ostaszewski2021,Anand2021}
Identifying important heuristics to scan the discrete neighbourhood
more selectively will  further improve the efficiency of DISCO-VQE.\cite{Schebarchov2013,Schebarchov2014}

Symmetry-preserving unitary product states 
offer key advantages for practical quantum algorithms:
they can parametrise an arbitrarily accurate wave function; they preserve the physical 
symmetries of the Hamiltonian at every truncation; 
and they have a low CNOT count suitable for noisy quantum hardware.
Conserving physical symmetries is essential
for creating suitable initial states for fault-tolerant quantum 
phase estimation, where efficiency depends on the overlap of the initial and exact states.
Furthermore, formulating the generalised fermionic operators as Lie algebra generators for 
unitary rotations moves beyond the many-body unitary coupled cluster framework. 
This new perspective fully exploits the natural capabilities of qubit rotations in quantum computation, 
bridging the gap between fermionic operators and hardware-efficient \ansatze{}.
Ultimately, these advances set a new standard for quantitative, physically accurate, 
and gate efficient quantum circuits for simulating strongly-correlated
chemical problems using near-term quantum computers.

\section{References and Notes}

\bibliographystyle{Science}
\bibliography{manuscript}

\begin{thebibliography}{10}

\bibitem{Bauer2020}
B.~Bauer, S.~Bravyi, M.~Motta, G.~K.-L. Chan, {Quantum Algorithms for Quantum
  Chemistry and Quantum Materials Science}, {\it Chem.\ Rev.\/} {\bf 120},
  12685 (2020).

\bibitem{AspuruGuzik2005}
A.~Aspuru-Guzik, A.~D. Dutoi, P.~J. Love, M.~Head-Gordon, {Simulated Quantum
  Computation of Molecular Energies}, {\it Science\/} {\bf 309}, 1704 (2005).

\bibitem{McArdle2020}
S.~McArdle, S.~Endo, A.~Aspuru-Guzik, S.~C. Benjamin, X.~Yuan, {Quantum
  computational chemistry}, {\it Reviews of Modern Physics\/} {\bf 92}, 15003
  (2020).

\bibitem{Peruzzo2014}
A.~Peruzzo, {\it et~al.\/}, {A variational eigenvalues solver on a photonic
  quantum processor}, {\it Nat.\ Comm.\/} {\bf 5}, 4213 (2014).

\bibitem{Kandala2017}
A.~Kandala, {\it et~al.\/}, {Hardware-efficient variational quantum eigensolver
  for small molecules and quantum magnets}, {\it Nature\/} {\bf 549}, 242
  (2017).

\bibitem{Anand2022}
A.~Anand, {\it et~al.\/}, {A quantum computing view on unitary coupled cluster
  theory}, {\it Chem.\ Soc.\ Rev.\/} {\bf 51}, 1659 (2022).

\bibitem{HelgakerBook}
T.~Helgaker, P.~J{\o}rgensen, J.~Olsen, {\it {Molecular Electronic-Structure
  Theory}\/} (John Wiley {\&} Sons, 2000).

\bibitem{Bartlett2007}
R.~J. Bartlett, M.~Musia{\l{}}, {Coupled-cluster theory in quantum chemistry},
  {\it Rev.\ Mod.\ Phys.\/} {\bf 79}, 291 (2007).

\bibitem{Evangelista2019}
F.~A. Evangelista, G.~K.-L. Chan, G.~E. Scuseria, {Exact parameterization of
  fermionic functions via unitary coupled cluster theory}, {\it J.\ Chem.\
  Phys.\/} {\bf 151}, 244122 (2019).

\bibitem{Izmaylov2020}
A.~F. Izmaylov, M.~D\'{i}az-Tinoco, R.~A. Lang, {On the order problem in
  construction of unitary operators for the variational quantum eigensolver},
  {\it Phys.\ Chem.\ Chem.\ Phys.\/} {\bf 22}, 12980 (2020).

\bibitem{Grimsley2020}
H.~R. Grimsley, D.~Claudino, S.~E. Economou, E.~Barnes, N.~J. Mayhall, {Is the
  Trotterized UCCSD Ansatz Chemically Well-Defined?}, {\it J.\ Chem.\ Theory
  Comput.\/} {\bf 16}, 1 (2020).

\bibitem{Tsuchimochi2020}
T.~Tsuchimochi, Y.~Mori, S.~L. Ten-no, {Spin-projection for quantum
  computation: A low-depth approach to strong correlation}, {\it Phys. Rev.
  Research\/} {\bf 2}, 043142 (2020).

\bibitem{Grimsley2019}
H.~R. Grimsley, S.~E. Economou, E.~Barnes, N.~J. Mayhall, {An adaptive
  variational algorithm for exact molecular simulations on a quantum computer},
  {\it Nat.\ Comm.\/} {\bf 10}, 3007 (2019).

\bibitem{Tang2021}
H.~L. Tang, {\it et~al.\/}, {Qubit-ADAPT-VQE: An Adaptive Algorithm for
  Constructing Hardware-Efficient Ans\"{a}tze on a Quantum Processor}, {\it PRX
  Quantum\/} {\bf 2}, 020310 (2021).

\bibitem{Chan2021}
H.~H.~S. Chan, N.~Fitzpatrick, J.~Segarra-Mart\'{i}, M.~J. Bearpark, D.~P. Tew,
  {Molecular excited state calculations with adaptive wavefunctions on a
  quantum eigensolver emulation: reducing circuit depth and separating spin
  states}, {\it Phys.\ Chem.\ Chem.\ Phys.\/} {\bf 23}, 26438 (2021).

\bibitem{Tsuchimochi2022}
T.~Tsuchimochi, M.~Taii, T.~Nishimaki, S.~L. Ten-no, {Adaptive construction of
  shallower quantum circuits with quantum spin projection for fermionic
  systems}  arxiv:2205.07097 (2022).

\bibitem{Yordanov2021}
Y.~S. Yordanov, V.~Armaos, C.~H.~W. Barnes, D.~R.~M. Arvidsson-Shukur,
  {Qubit-excitation-based adaptive variational quantum eigensolver}, {\it
  Commun.\ Phys.\/} {\bf 4}, 228 (2021).

\bibitem{Shkolnikov2021}
V.~O. Shkolnikov, N.~J. Mayhall, S.~E. Economou, E.~Barnes, {Avoiding symmetry
  roadblocks and minimizing the measurement overhead of adaptive variational
  quantum eigensolvers}  arxiv:2109.05340 (2021).

\bibitem{Rubin2021}
N.~C. Rubin, J.~Lee, R.~Babbush, {Compressing Many-Body Fermion Operators Under
  Unitary Constraints}  (2021).

\bibitem{Jordan1928}
P.~Jordan, E.~Wigner, {\"{U}ber das Paulische \"{A}quivalenzverbot}, {\it Z.\
  Phys.\/} {\bf 47}, 631 (1928).

\bibitem{Bravyi2002}
S.~B. Bravyi, A.~Y. Kitaev, {Fermionic Quantum Computation}, {\it Ann.\
  Phys.\/} {\bf 298}, 210 (2002).

\bibitem{GilmoreLieBook}
R.~Gilmore, {\it {Lie Groups, Physics, and Geometry: An Introduction for
  Physicists, Engineers, and Chemists}\/} ({Dover Publications Inc.}, 2008),
  first edn.

\bibitem{HallBook}
B.~C. Hall, {\it Lie Groups, Lie Algebras, and Representations\/} ({Springer
  Chem}, 2015).

\bibitem{Lee2019c}
J.~Lee, W.~J. Huggins, M.~Head-Gordon, K.~B. Whaley, {Generalized Unitary
  Coupled Cluster Wave functions for Quantum Computation}, {\it J.\ Chem.\
  Theory Comput.\/} {\bf 15}, 311 (2019).

\bibitem{Grimsley2022}
H.~R. Grimsley, G.~S. Barron, E.~Barnes, S.~E. Economou, N.~J. Mayhall,
  {ADAPT-VQE is insensitive to rough parameter landscapes and barren plateaus}
  arxiv:2204.07179 (2022).

\bibitem{Schebarchov2013}
D.~Schebarchov, D.~J. Wales, {Communication: A new paradigm for structure
  prediction in multicomponent systems}, {\it J.\ Chem.\ Phys.\/} {\bf 139},
  221101 (2013).

\bibitem{Schebarchov2014}
D.~Schebarchov, D.~J. Wales, {Structure Prediction for Multicomponent Materials
  Using Biminima}, {\it Phys.\ Rev.\ Lett.\/} {\bf 113}, 156102 (2014).

\bibitem{Roeder2018}
K.~R{\"{o}}der, D.~J. Wales, {Mutational Basin-Hopping: Combined Structure and
  Sequence Optimization for Biomolecules}, {\it J.\ Phys.\ Chem.\ Lett.\/} {\bf
  9}, 6169 (2018).

\bibitem{lis87}
Z.~Li, H.~A. Scheraga, Monte carlo-minimization approach to the multiple-minima
  problem in protein folding, {\it PNAS\/} {\bf 84}, 6611 (1987).

\bibitem{Wales1997}
D.~J. Wales, J.~P.~K. Doye, {Global Optimization by Basin-Hopping and the
  Lowest Energy Structures of Lennard-Jones Clusters Containing up to 110
  Atoms}, {\it J.\ Phys.\ Chem.\ A\/} {\bf 101}, 5111 (1997).

\bibitem{Hehre1969}
W.~J. Hehre, R.~F. Stewart, J.~A. Pople, {Self-Consistent Molecular-Orbital
  Methods. I. Use of Gaussian Expansions of Slater-Type Atomic Orbitals}, {\it
  J.\ Chem.\ Phys.\/} {\bf 51}, 2657 (1969).

\bibitem{Matsuzawa2020}
Y.~Matsuzawa, Y.~Kurashige, {Jastrow-type Decomposition in Quantum Chemistry
  for Low-Depth Quantum Circuits}, {\it J.\ Chem.\ Theory Comput.\/} {\bf 16},
  944 (2020).

\bibitem{Note1}
The non-parallelity error is defined as the difference between the maximum and
  minimum error along a binding curve.

\bibitem{Sapova2021}
M.~D. Sapova, A.~K. Fedorov, {Variational quantum eigensolver techniques for
  simulating carbon monoxide oxidation}  arxiv:2108.111.67 (2021).

\bibitem{Sim2021}
S.~Sim, J.~Romero, J.~F. Gonthier, A.~A. Kunitsa, {Adaptive pruning-based
  optimization of parametrized quantum circuits}, {\it Quantum Sci.\
  Technol.\/} {\bf 6}, 025019 (2021).

\bibitem{Yordanov2020}
Y.~S. Yordanov, D.~R.~M. Arvidsson-Shukur, C.~H.~W. Barnes, {Efficient quantum
  circuits for quantum computational chemistry}, {\it Phys.\ Rev.\ A\/} {\bf
  102}, 062612 (2020).

\bibitem{LeBlanc2015}
J.~P.~F. LeBlanc, {\it et~al.\/}, {Solutions of the Two-Dimensional Hubbard
  Model: Benchmarks and Results from a Wide Range of Numerical Algorithms},
  {\it Phys.\ Rev.\ X\/} {\bf 5}, 041041 (2015).

\bibitem{Stokes2020}
J.~Stokes, J.~Izaac, N.~Killoran, G.~Carleo, {Quantum natural gradient}, {\it
  Quantum\/} {\bf 4}, 269 (2020).

\bibitem{Koczor2022}
B.~Koczor, S.~C. Benjamin, {Quantum analytic descent}, {\it Phys.\ Rev.\
  Research\/} {\bf 4}, 023017 (2022).

\bibitem{Rattew2019}
A.~G. Rattew, S.~Hu, M.~Pistoia, R.~Chen, S.~Wood, {A Domain-agnostic,
  Noise-resistant, Hardware-efficient, Evolutional Variational Quantum
  Eigensolver}  arXiv:1910.09694 (2019).

\bibitem{Chivilikhin2020}
D.~Chivilikhin, {\it et~al.\/}, {MoG-VQE: Multiobjective genetic variational
  quantum eigensolver}  arxiv:2007.04424 (2020).

\bibitem{Zhang2021}
S.-X. Zhang, C.-Y. Hsieh, S.~Zhang, H.~Yao, {Differentiable Quantum
  Architecture Search}  arxiv:2010.08561 (2021).

\bibitem{Ostaszewski2021}
M.~O.~E. Grant, M.~Benedetti, {Structure optimization for parameterized quantum
  circuits}, {\it Quantum\/} {\bf 5}, 391 (2021).

\bibitem{Anand2021}
A.~Anand, M.~Degroote, A.~Aspuru-Guzik, {Natural evolutionary strategies for
  variational quantum computation}, {\it Mach.\ Learn.: Sci.\ Technol.\/} {\bf
  2}, 045012 (2021).

\end{thebibliography}

\section{Acknowledgements}
\noindent
The authors gratefully acknowledge the use of the 
University of Oxford Advanced Research Computing (ARC) 
facility in carrying out this work. 
\href{http://dx.doi.org/10.5281/zenodo.22558}{http://dx.doi.org/10.5281/zenodo.22558}

\noindent\textbf{Funding:}\\
H.G.A.B.\ was supported by New College, Oxford through the Astor Junior Research Fellowship.
D.M.D. was supported by the EPSRC Hub in Quantum Computing and Simulation (EP/T001062/1).
D.J.W.\ gratefully acknowledges financial support from the EPSRC.

\noindent\textbf{Author contributions:}\\
H.G.A.B., D.P.T., and D.J.W.\ conceived the project.
H.G.A.B., D.M.D., and D.J.W.\ designed the discrete optimisation algorithm.
D.M.D., H.G.A.B., and D.P.T.\ designed and analysed the s-UPS ansatz.
H.G.A.B.\ wrote the DISCO-VQE code and carried out the simulations.  
H.G.A.B.\ and D.M.D. performed the data analysis.
All authors contributed to writing the manuscript.
H.G.A.B.\ initiated and managed the collaboration.

\noindent\textbf{Competing interests:}\\
The authors declare that they have no competing interests.

\noindent\textbf{Data and materials availability:}\\
The data required to reproduce figures are hosted at \href{https://dx.doi.org/10.5281/zenodo.6784056}{https://dx.doi.org/10.5281/zenodo.6784056}.
All other data required to evaluate the conclusions are present in the paper or the supplementary materials.

\section{Supplementary Materials}
\noindent
Supplementary Text\\
Figs.\ S1 to S3\\
Table S1\\
References (45--58)

\end{document}